\documentclass[english]{lni}

\usepackage[utf8]{inputenc}
\usepackage[english]{babel}

\usepackage{comment}

\usepackage{booktabs}

\usepackage[]{blindtext}

\usepackage{amsmath,amssymb,amsfonts}
\usepackage{algorithmic}
\usepackage{graphicx}
\usepackage{textcomp}
\usepackage{xcolor}
\usepackage[braket, qm]{qcircuit}
\usepackage[sorting=none, style=nature]{biblatex}
\def\BibTeX{{\rm B\kern-.05em{\sc i\kern-.025em b}\kern-.08em
    T\kern-.1667em\lower.7ex\hbox{E}\kern-.125emX}}

\usepackage{todonotes}

\begin{document}
\title[QV paper]{A comparison of HPC-based quantum computing simulators using Quantum Volume}
\author[2]{Lourens van Niekerk}{lourens.van-niekerk@gwdg.de}{}
\author[2]{Dhiraj Kumar}{dhiraj.kumar@gwdg.de}{}
\author[2]{Aasish Kumar Sharma}{aasish-kumar.sharma@gwdg.de}{}
\author[1]{Tino Meisel}{tino.meisel@gwdg.de}{}%
\author[1]{Martin Leandro Paleico}{martin-leandro.paleico@gwdg.de}{}%
\author[1]{Christian Boehme}{christian.boehme@gwdg.de}{}

\affil[1]{GWDG\\Göttingen\\Germany}
\affil[2]{Göttingen University\\Göttingen\\Germany}
\maketitle

\begin{abstract}
This paper compares quantum computing simulators running on a single CPU or GPU-based HPC node using the Quantum Volume benchmark commonly proposed for comparing NISQ systems. As simulators do not suffer from noise, the metric used in the comparison is the time required to simulate a set Quantum Volume. The results are important to estimate the feasibility of proof of concept studies and debugging of quantum algorithms on HPC systems. Besides benchmarks of some commonly used simulators, this paper also offers an overview of their main features, a review of the state of quantum computing simulation and quantum computing benchmarking, and some insight into the theory of Quantum Volume.
\end{abstract}


\section{Introduction}

Understanding and utilizing the power of quantum computations, as compared to classical computations, is the ultimate mission behind the accelerating research and funding done in quantum computing. In the words of John Preskill: "Can we control complex quantum systems and if we can, so what?" \cite{Preskill2012}. In the last decade, the quantity of quantum physics papers, of which many involve quantum computing, that have been published on the \textit{ArXiv} online archive, has effectively doubled from 4356 articles in 2014, to 8616 articles in 2023 \cite{Arxiv-quant-ph}. We are still, however, in the Noisy Intermediate Scale Quantum (NISQ) era \cite{Preskill2018}, and access to high-quality quantum computers is still rare, restricted and expensive. This is the primary reason why quantum (computing) simulators \cite{Johnson2014} are currently, despite a lack of sufficient memory for large quantum circuit simulation, key to the development of applications which show promise for quantum supremacy (faster results, more accuracy, more features, etc.) \cite{Preskill2012} once quantum computers can sustain minimal errors for large quantum circuits. 

The dissemination of quantum simulators, such as Qiskit \cite{Qiskit}, Cirq \cite{cirq}, and Qulacs \cite{qulacs}, naturally raises a question: what distinguishes one simulator from another? Benchmarking provides a crucial tool for understanding and characterizing these simulators, as it compares their underlying mathematical formalism and software implementation capabilities, which impose stringent limits on the system sizes that can be explored. Note that while simulators mimic a quantum system's behavior, emulators replicate the actual quantum hardware, albeit classically. In the context of quantum computing, simulators are often the focus, as they enable the exploration of larger system sizes and various quantum algorithms.

\subsection{Related Work}

Numerous propositions of benchmarks and metrics have been proposed for comparing different quantum computing (QC) systems \cite{Wang2022}. These benchmarks can be divided into three categories: Physical Benchmarks, Aggregated Benchmarks, and Application-level Benchmarks. Physical Benchmarks are only concerned with the hardware of quantum computers with desirable results such as high qubit quantities, high connectivity (e.g., all-to-all connected), and high gate fidelities. Aggregated benchmarks present more of a comprehensive and holistic assessment of a quantum processor. The most popular such benchmark is quantum volume (QV) \cite{Cross2018}, but other proposals, such as circuit-layer operations per second (CLOPS) \cite{Wack2021}, have been proposed as an attempt to address the short-comings of QV. Proposals for application-level benchmarks are, amongst others, Quantum LINPACK \cite{Dong2021}, QPack \cite{Mesman2021}, and QASMBench \cite{Li2023}. 

Some of these benchmarks, including custom benchmarks, have also been used to measure the efficiency of simulators. In 2020, Luo et al. \cite{Luo2019}, the developers of a Julia-based quantum simulator called \textit{Yao}, benchmarked various simulators investigated in this paper, as well as other simulators, by comparing the execution of a single gate on various circuit sizes. They implemented well-established single-qubit gates such as X, H, and T gate, as well as the two-qubit CNOT gate, on the second qubit of a quantum circuit ranging from 4 to 25 qubits, and found that the Qulacs simulator was either superior or equivalent to the Yao simulator in terms of speed, particularly when graphics processing unit (GPU) acceleration was utilized. 

The team of developers of the Qulacs simulator \cite{qulacs} in 2021 utilized randomized benchmarking by constructing a 'rotation layer' and 'CNOT layer' ten times in alternating order. They found that Qulacs, with their circuit optimization method, generally outperformed the rest of their listed simulators in terms of speed for this benchmark. 

In 2023, Strano et al. \cite{Strano2023} released the Qrack simulator and benchmarked it compared to other simulators concerning the Quantum Fourier Transform (QFT) on a single GPU. Qrack outperformed the other selected simulators by a significant margin when all simulators were initialized to the zero-state. However, it only outperformed most other simulators when they were initialized to the GHZ state \cite{Greenberger2007}. 

Most recently, in early 2024, Jamadagni et al. \cite{Jamadagni2024} did an extensive and detailed benchmarking study of many quantum simulators available. They benchmarked the simulators by speed of operation when performing Heisenberg dynamics, randomized benchmarking, and QFTs for quantum circuits with eight and up to 32 qubits. These benchmarks were tested on a single node with either single-threading, multi-threading, or GPU acceleration. Their findings broadly show that Qiskit is the top performer, particularly at high qubit counts, though Qiskit could be accelerated even further when integrated with the cuquantum  software development kit (SDK) by Nvidia \cite{Bayraktar2023}.

\section{Benchmarking Quantum Computing Simulators}

\subsection{What is quantum volume?} 

For benchmarking quantum simulators, we chose to use quantum volume considering how widely adopted it currently is by the quantum computing community, and since the majority of the drawbacks raised against QV as a benchmark do not apply to quantum simulators as they do to quantum processing units (QPUs). Physical benchmarks do not concern simulators run on high-performance computing (HPC) systems since all operations are assumed to have perfect fidelity and all-to-all connectivity, with the quantity of qubits they are able to simulate limited only by the random access memory (RAM) of the system. Available HPC system RAM of a single node is typically enough to handle a quantum statevector with a maximum qubit count roughly in the low thirties. Each incremental qubit to the statevector doubles the memory, and potentially requires double the amount of nodes to simulate. 

Quantum volume, as defined by \textit{Cross et al.} \cite{Cross2018}, is a single metric $V_Q$ representing the maximum size circuit consisting of $n$-qubits (the width of a circuit) and $n$-layers (the depth of a circuit) that can be run on a quantum computer with a heavy output probability of at least $66.\overline{6}\%$, or $2/3$. A \textit{layer} consists of a random permutation of the order of all qubits within the statevector, followed by a row of Haar-random $SU(4)$ unitary gates, where 'Haar-random' refers to random gates according to the Haar measure, and $SU(4)$ refers to $4 \times 4$ special unitary gates (with a determinant of $1$). After all layers are applied, measurements are taken of all qubits. The \textit{heavy output} \cite{Aaronson2016} of a statevector is the set of measurements of a statevector that are more probable than the measurement with the median probability. Accordingly, the \textit{heavy output probability} (HOP) is the probability of achieving a measurement in the heavy output of a statevector. If a sufficient error is introduced into any QV test, e.g., decoherence and gate error, the HOP would converge to 50\% as the number of gates increases in a circuit, i.e., with each qubit incremented in a QV test. If, however, a QV test is errorless, then the HOP would converge to $(1 + ln2)/2$, which is roughly $84.66\%$ (proof presented in Section 2.2).
QV's beauty is that it practically combines all physical benchmark metrics into a single metric that represents quality and scale. We showcase the scaling of the QV in Fig. 1
below.

Our goal is to test for the speed at which current quantum simulators can perform QV tests on a single node (with/without GPU), assuming all-to-all connectivity and ideal circuits where no error is introduced. 

All frameworks presented are considered to perform the intended calculations correctly, so the question of computation speed is the most relevant for simulations. The general trend that simulation time increases per qubit increment by slightly more than double (twice as large statevector plus additional overhead) means that a faster simulator would sustainably reduce compute time, which allows for more research and lowers resource costs.  

\begin{figure}[h]
    \centering
    \begin{minipage}[t]{0.15\textwidth}
    \Qcircuit @C=1em @R=1.6em {
 \lstick{\ket{0}} & \multigate{3}{\#} & \multigate{1}{\mu} & \multigate{3}{\#} & \multigate{1}{\mu} & \multigate{3}{\#} & \multigate{1}{\mu} & \multigate{3}{\#} & \multigate{1}{\mu} & \meter & \qw \\
 \lstick{\ket{0}} & \ghost{\#} & \ghost{\mu} & \ghost{\#} & \ghost{\mu}& \ghost{\#} & \ghost{\mu} & \ghost{\#} & \ghost{\mu} & \meter & \qw \\
 \lstick{\ket{0}} & \ghost{\#} & \multigate{1}{\mu} & \ghost{\#} & \multigate{1}{\mu} & \ghost{\#} & \multigate{1}{\mu} & \ghost{\#} & \multigate{1}{\mu} & \meter & \qw \\
 \lstick{\ket{0}} & \ghost{\#} & \ghost{\mu} & \ghost{\#} & \ghost{\mu} & \ghost{\#} & \ghost{\mu} & \ghost{\#} & \ghost{\mu} & \meter & \qw 
 \gategroup{1}{2}{4}{3}{.7em}{--} \gategroup{1}{4}{4}{5}{.7em}{--} \gategroup{1}{6}{4}{7}{.7em}{--} \gategroup{1}{8}{4}{9}{.7em}{--} 
}
\end{minipage}
\vspace{0.2cm}
\caption{Quantum volume test with 4 qubits and 4 layers (dashed blocks). The \# gate represents a random permutation of the qubit amplitudes achieved via SWAP gates, and the $\mu$ gate is a Haar-random $SU(4)$ unitary gate.}
\label{qv4}
\end{figure}
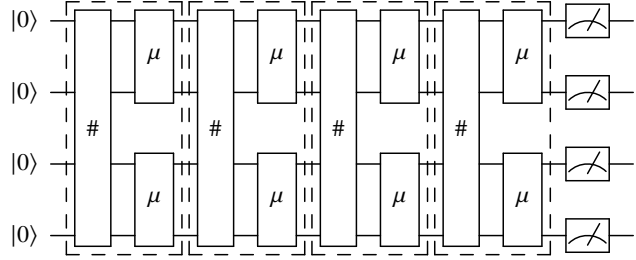

\subsection{HOP Convergence}

It is crucial to check that a noiseless QV test produces the desired HOP convergence as a sanity check that the test is implemented correctly. Another sanity check is that the expected number of gates in the QV circuit for specific numbers of qubits is the same across all simulators. Here, we provide proof for HOP convergence in a noiseless QV test, which seems worthwhile since we could not find a straightforward mathematical proof in the literature for the convergence value as $(1 + ln2) /2 $. Consider that the state of an $n$-qubit quantum register after applying a QV circuit (or at any point in the circuit) can be described by its $N \times N$ (hermitian) density matrix $\rho$ \cite{Scherer2019}, where $N = n^2$. Since the QV circuit consists of layers of random permutations and Haar-random $SU(4)$ unitary gates, we could consider the entire circuit as a single Haar-random $SU(N)$ unitary gate and thus consider the density matrix in light of the Gaussian Unitary Ensemble (GUE). If we assume matrices $A,B,C$ are such that $A,B \in SU(2^r)$ and $C\in SU(2^s)$, with all three Haar-random unitary gates, then it is still the case that the matrix multiplication $AB \in SU(2^r)$ is a Haar-random unitary gate, and the dot product $A \cdot B \in SU(2^{r+s})$ is a Haar-random unitary gate. Considering the random permutations are simply a random reordering of the amplitudes in the quantum statevector, which equates to a random reordering of the (random) matrix values in the density matrix, then the dot products of the Haar-random $SU(4)$ unitary gates, and matrix multiplications, amount to a single Haar-random $SU(N)$ unitary gate. 

The joint probability density function (jpdf) of all eigenvectors, $P(\mathbf{e})$, of $\rho$ is
\begin{equation}
    P(\mathbf{e}) = C_N \delta\left( 1 - \sum^N_{j=1} |e_j|^2 \right) \;\;\;\;\;\;\;\;\; ; \;\;\;\;\;\;\;\;\; \delta(x) = \lim_{\epsilon \rightarrow 0^+} \frac{1}{2\sqrt{\pi \epsilon}}exp\left({\dfrac{-x^2}{4\epsilon}}\right)
\end{equation}
where $Z_N$ and $C_N$ are normalization constants, $e_j$ are eigenvectors with $\mathbf{e}$ defining the set of eigenvectors, and
$\delta(x)$ is the Dirac delta distribution \cite{Livan2017}. This behavior is due to $\rho$ belonging to the GUE since $\rho$ is hermitian, has eigenvectors each with a norm of $1$, and has complex values at all entries except the main diagonal. Both these jpdfs are properties of the GUE. 
The eigenvectors of $\rho$ correspond to a quasi-probability distribution of a specific measurement outcome. Thus, the area above the median of the marginal distribution of Equation (2), i.e., the marginal pdf, $P_{mar}(y)$, of a single eigenvector in $\rho$, would provide insight into the HOP.
According to \textit{Livan et al.}, the marginal pdf of the norm of any eigenvector $e_j$ of $\rho$ would be 
\begin{equation}
    P_{mar}(y) = \int d^2e_1\cdots d^2e_N\delta(y - |e_j|^2)P(\mathbf{e}) = (N-1)(1-y)^{N-2}
\end{equation}
for $0 \leq y \leq 1$, where $y$ represents the absolute value of the determinant of an arbitrary eigenvector. We need to know what the marginal pdf converges to. Thus, if we choose $\eta = y N$ and take $N \rightarrow \infty$, then we get
\begin{equation}
    P_{mar}(\eta) = \lim_{N \rightarrow \infty} \frac{1}{N}P_{mar}\left( \frac{\eta}{N} \right) = e^{-\eta},
\end{equation}
which is simply the exponential distribution $Exp(\lambda)$ with $\lambda=1$, though some authors refer to it as the \textit{Porter-Thomas} distribution \cite{Porter1956}. The median for $Exp(1)$ is $ln(2)$. Calculating the amount of eigenvector amplitudes above $ln(2)$ for $N \rightarrow \infty$, we get 
\begin{equation}
    \int_{ln(2)}^{\infty} xe^{-x}dx = \frac{1 + ln(2)}{2} \approx 0.846574.
\end{equation}
Given correct quantum volume implementation, the heavy output probability should converge to $\approx 84.6\%$ as the number of qubits increases \cite{Aaronson2016}.

\section{Investigated Frameworks}

Our interest lies in simulating ideal circuits on CPUs and GPUs at optimal speeds, for various frameworks. In this section, we describe each framework that was benchmarked using QV. We chose these frameworks based on whether they \textit{(a)} have proven fast execution times regarding the methods mentioned in the Section 1.1, \textit{(b)} are open-source, \textit{(c)} widely cited in related papers, and \textit{(d)} are still active (regular and recent updates). 

\textbf{Qiskit Aer} \cite{Qiskit} is a high-performance simulation toolkit in the IBM Quantum ecosystem. It uses OpenQASM 3.0 \cite{openqasm3}, the industry standard intermediate representation (IR), as its IR. OpenQASM is often referred to simply as QASM (quantum assembly language). Qiskit Aer acts on top of the Qiskit Terra base package. Qiskit Aer offers various simulation methods, such as statevector, density matrix, and tensor\_network, with statevector most suitable to our purposes. Qiskit has a QuantumVolume function, aligned with the original QV paper \cite{Cross2018} when setting \textit{classical\_permutation=False}. Qiskit supports Message Passing Interface (MPI) for efficient multinode scaling.

\textbf{Qulacs} \cite{qulacs} is a library developed by QunaSys, Osaka University, NTT Computer and Data Science Laboratories, and Fujitsu. It employs a hybrid Schrödinger-Feynman method \cite{Markov2018} for versatile simulation between Schrödinger equations and Feynman paths. Due to Qulacs lacking a quantum volume function, our team manually coded a QV function in Python, following the QV paper \cite{Cross2018}. Qulacs allows GPU acceleration, and a fork called mpiQulacs \cite{Imamura2022} utilizes MPI for efficient multinodal scaling, though with restricted method support. 

\textbf{Cirq} \cite{cirq}, a Google Quantum AI Python library, models quantum programs with circuits, consisting of Moments. Moments encapsulate operations on the same time slice. Cirq offers pure state and mixed state simulations for small circuit testing, with numerous algorithms and experiments pre-implemented in the standard code

\textbf{Qsim} \cite{qsim} is developed at Google Quantum AI for quantum circuit simulation. It is a full wave function simulator written in C++ with a Python wrapper. Qsim allows researchers to test quantum algorithms before running them on actual quantum hardware. It uses multithreading and gate fusion to accelerate the simulations. Qsim is an extension of Cirq as quantum circuits are designed using Cirq.

\textbf{Intel Quantum SDK} \cite{Wu2023} is a C++-based quantum development suite that supports hybrid quantum-classical algorithm creation, providing various features like backends, state vector setups, noise models, and compiler routines. It runs on CPUs with MPI parallelization, but its C++ foundation can be challenging to learn. A Python interface is available, but it doesn't support parallelization with MPI. For our benchmarking, we exported QASM code from Qiskit, imported it using the SDK's QASM translator, and ran it from Intel's publicly provided Docker container. Note that the SDK only parallelizes across a power of 2 threads, so we used a maximum of 64 cores.

\textbf{Qrack} \cite{Strano2023} is an open source full-stack QC simulator library for HPC systems written in C++11. It has an external-dependency-free CPU and GPU simulator engine that depends only on OpenCL. It also supports NVIDIA's Compute Unified Device Architecture (CUDA) and other options for parallel computing. Qrack includes support for the Qiskit\_Aer QSAM simulator and \textit{pyqrack} Python wrappers to support the developers. 

\textbf{Qibo} \cite{qibo_paper} is an open-source QC framework for quantum circuit simulation and optimization. Its variant, qibojit \cite{qibojit_paper}, offers a JIT compiler with custom numba kernel for CPU multi-threading and cupy/cuQuantum-based multi-GPU acceleration, distributing states across devices for faster execution. Unlike Qibo's CPU-only simulations, qibojit's hybrid approach significantly boosts performance for large-scale quantum computations, making it our preferred choice for benchmarking.

The package versions used for each simulator were qiskit 0.45.3, qiskit-terra 0.25.0, qiskit-aer 0.12.2, qiskit-aer-gpu 0.11.2, qulacs 0.6.1, qulacs-gpu 0.3.1, cirq 1.3.0, qsimcirq 0.16.3, intel quantum sdk 1.0, pyqrack 1.28.0, qibo 0.2.7, qibojit 0.1.3.

\section{Performance Evaluation}

\subsection{Hardware and software}
The exact specifications are listed in Table 1.
Specific simulators, such as Qiskit-Aer and Qulacs, support multi-node parallelization via MPI.

\begin{table}
    \centering
    \begin{tabular}{lll}
        \toprule
        Job type & CPU & CPU with GPU accelerator \\
        \midrule
        HPC system & Emmy & Grete \\  
        CPUs & 2 Intel Cascadelake 9242 & 2 AMD Zen3 EPYC 7513 \\
        CPU cores & 96 & 64 \\
        RAM & 384 GB & 512 GB \\
        GPUs & & 4 Nvidia Tesla A100 40GB \\
        \bottomrule
    \end{tabular}
    \caption{Hardware specifications for single node jobs}
    \label{tab:hardware}
\end{table}

\subsection{Execution}

All jobs were executed on only a single node. A 12-hour time limit was set for simulations, with 100 shots used for small qubit quantities. For larger quantities with slower execution times, 10 or 1 shot was used instead. Simulators were verified to converge to $\sim 84.6\%$ HOP, except for Cirq and Qsim, which showed inconsistent convergence patterns for even and odd qubit numbers, oscillating between $\sim 84.6\%$ and $\sim 54\%$. This is unclear why. Simulations used all-to-all connectivity, thus omitting additional SWAP gates besides the permutation SWAP gates. Execution times were calculated solely for circuit execution, including circuit optimization if applied. Thus circuit construction and HOP calculation were excluded.

\subsection{Results}

\begin{figure}[ht]
    \centering
    \includegraphics[width=12cm,height=6cm]{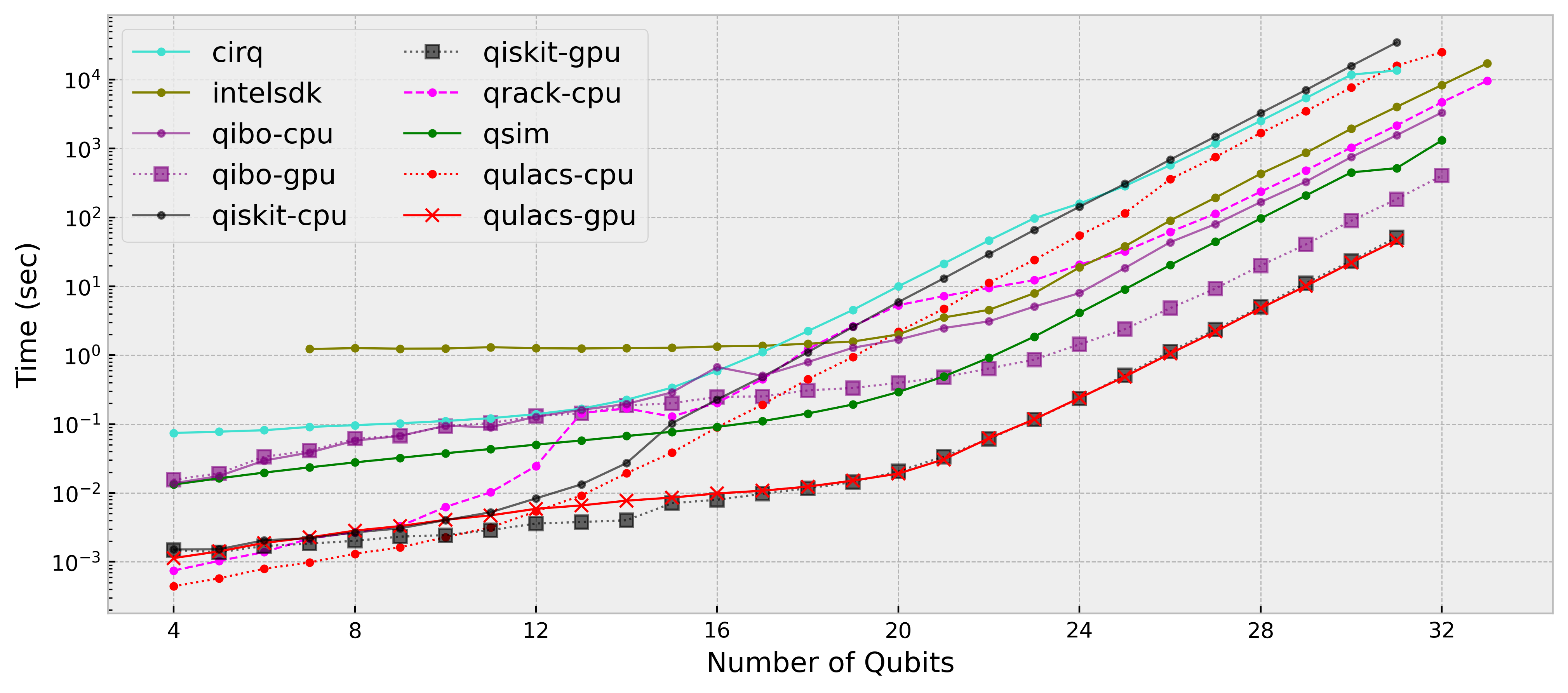}
    \caption{Logarithmic times in seconds for simulating a single QV circuit of the given qubit width (equaling the circuit depth) for tested simulator frameworks on a single node. }
    \label{fig:combined_result}
\end{figure}

The benchmarking results in Figure 2
unequivocally show that GPU-accelerated circuits, particularly those of Qulacs and Qiskit, offer the best performance, with Qiskit-GPU the fastest up to 16 qubits. Qulacs CPU implementations excel for small circuits, but GPU acceleration provides the fastest execution times for larger circuits. In CPU-only execution, Qulacs is the fastest for circuits smaller than 16 qubits, and qsim the fastest for larger circuits.

Intel Quantum SDK performs well for larger qubit counts, still only ranking fourth among CPU-exclusive codes, only utilizing 64 out of 96 available cores due to MPI constraints. It fares worse for low qubit counts, potentially due to overhead from running in a container and not saturating compute units. Notably, only Intel SDK and Qrack CPU could simulate 33 qubits on the available nodes, while other simulators ran out of memory at 32 qubits. Qrack performed slightly better than Intel SDK.

The qibojit framework has a performance that is in the midfield of all benchmarks. Both CPU and GPU accelerated runs execute QVs with 32 qubits before running out of memory. While its multi-threaded CPU execution times are mediocre for smaller QVs, qibojit shows good scalability from 20 qubits. It becomes the second-fastest application with CPU only after qsim for larger QVs. However, its GPU-accelerated performance is surprisingly poor, resembling CPU-only execution times up to 14 qubits and only showing better performance afterward. Compared to other benchmarked GPU-accelerated QVs, qibojit has the highest execution times.

Qsim performed better than all other CPU based frameworks, 
but still worse than all GPU based frameworks. Qsim was fast enough to run 32 qubits successfully within the 12-hour time constraint. Cirq, on the other hand, was the worst-performing framework for low qubit counts but surpassed Qulacs CPU for high qubit counts. 

Based on our benchmarks, we recommend the following (see Table 2).

\begin{table}
\centering
\begin{tabular}{llp{4cm}}
\toprule
Circuit Size & Recommended Frameworks & Notes \\
\midrule
$<$ 16 qubits & Qiskit CPU, Qulacs CPU, Qsim & Qulacs CPU is fastest \\
16-24 qubits & Qiskit GPU, Qulacs GPU, Qsim, (qibojit) & Qibojit offers seamless GPU and CPU execution, but with a performance tradeoff.  \\
24-32 qubits & Qiskit GPU, Qulacs GPU, Qsim, Qrack CPU & GPU acceleration recommended \\
$>$ 32 qubits & Intel Quantum SDK, Qrack CPU & Only frameworks that can simulate 33 qubits in our setup \\
\bottomrule
\end{tabular}
\caption{Recommendations for Quantum Circuit Simulation Frameworks}
\label{tab:recommendations}
\end{table}

\section{Conclusion and Outlook}

Our benchmarks show that GPU-based quantum computing simulations outperform CPU-based simulations in terms of time to solution. The cost per simulation is also lower, with the seven times higher cost of GPU nodes more than offset by more than ten times higher performance. Notably, GPU simulators show minimal performance differences, whereas CPU simulators exhibit significant variations, possibly due to inefficient shared memory parallelization. 
Currently, no simulator can handle more than 33 qubits on the hardware used; larger circuits require distributed memory parallelization across multiple nodes, which may favor CPU-based simulations. Future work will investigate these findings.

\section*{Acknowledgements}

The authors thank Dr. Niklas Bölter for his guidance and review. They also gratefully acknowledge the computing time made available to them on the high-performance computers Emmy and Grete at the NHR Center NHR-Nord@Göttingen. This center is jointly supported by the Federal Ministry of Education and Research and the state governments participating in the NHR (www.nhr-verein.de/unsere-partner).

\printbibliography

\end{document}